\documentclass[aps,prd,twocolumn,superscriptaddress,showpacs]{revtex4}
\usepackage{float,graphicx}

\begin{document}


\title{Constraints on $f(R)$ Cosmology in the Palatini Formalism}



\author{Baojiu~Li}
\email[Email address: ]{B.Li@damtp.cam.ac.uk}
\affiliation{Department of Physics \& Institute of Theoretical
Physics, the Chinese University of Hong Kong, Hong Kong SAR,
China}
\affiliation{Department of Applied Mathematics \&
Theoretical Physics, Centre for Mathematical Sciences, University
of Cambridge, Cambridge CB3 0WA, United Kingdom}

\author{K.~C.~Chan}
\email[Email address: ]{kcchan@phy.cuhk.edu.hk}
\affiliation{Department of Physics \& Institute of Theoretical
Physics, the Chinese University of Hong Kong, Hong Kong SAR,
China}

\author{M.-C.~Chu}
\email[Email address: ]{mcchu@phy.cuhk.edu.hk}
\affiliation{Department of Physics \& Institute of Theoretical
Physics, the Chinese University of Hong Kong, Hong Kong SAR,
China}


\date{\today}

\begin{abstract}
In this work we derive the covariant and gauge invariant
perturbation equations in general theories of $f(R)$ gravity in
the Palatini formalism to linear order and calculate the cosmic
microwave background (CMB) and matter power spectra for an
extensively discussed model, $f(R) = R + \alpha(-R)^\beta$, which
is a possible candidate for the late-time cosmic accelerating
expansion found recently. These spectra are discussed and found to
be sensitively dependent on the value of $\beta$. We are thus able
to make stringent constraints on $\beta$ from cosmological data on
CMB and matter power spectra: The three-year Wilkinson Microwave
Anisotropy Probe (WMAP) data alone gives a constraint $|\beta|
\lesssim \mathcal{O}(10^{{-3}})$ while the joint WMAP, Supernova
Lagacy Survey (SNLS) and Sloan Digital Sky Survey (SDSS) data sets
tightens this to $\beta \sim \mathcal{O}(10^{{-6}})$, about an
order of magnitude more stringent than the constraint from SDSS
data alone, which makes this model practically indistinguishable
from the standard $\Lambda\mathrm{CDM}$ paradigm.
\end{abstract}


\maketitle


\section{INTRODUCTION}

\label{sect:Intro} It is observed that the universe is now
undergoing accelerating expansion \cite{Riess1, Perlmutter,
Riess2}, which is also consistent with the three-year Wilkinson
Microwave Anisotropy Probe (WMAP) data \cite{WMAP3yr} and several
other cosmological observations. The usual ``explanation'' for
this involves a mysterious component, called the dark energy,
which drives this accelerating expansion. However, this dark
energy problem could also be attacked by modifying the theory of
gravity so that it departs from the standard general relativity
(GR) when the spacetime curvature becomes small. In one type of
modified gravity theories, the Ricci scalar $R$ in the
Einstein-Hilbert action is simply replaced by a function of $R$,
commonly known as $f(R)$. Indeed, in \cite{Carroll,Capozziello},
it was shown that by adding correction terms, such as $R^2$, $R^{a
b} R_{a b}$ and $R^{abcd}R_{abcd}$, to the action, the late time
accelerating cosmic expansion could be reproduced (see also
\cite{Nojiri2006} and references therein for related works).
Another argument in favor of such generalizations is that the
effective Lagrangian of the gravitational field generally will
include higher order terms in the curvature invariants as a result
of quantum corrections (see, \emph{e.g.}, \cite{QFTcurved}).

However, the conventional metric approach to $f(R)$ gravity leads
to fourth order equations which may exhibit violent instabilities
when matter is present in the weak gravity regime \cite{Dolgov}
(see however \cite{Nijiri2003} for a discussion). On the other
hand, in the Palatini variation of the action where the metric and
connection are treated as independent dynamical variables
\cite{Vollick}, the resultant equations are second order, which
are more tractable and concordant with field equations in other
branches of physics. In particular, the typical form $f(R) = R +
\alpha (-R)^\beta$ has been discussed extensively in the
literature as an alternative dark energy model which fits rather
well with the supernovae (SNe) Ia data, and it is also tested
using cosmic microwave background (CMB) shift parameter and baryon
acoustic oscillation (BAO) in \cite{Cap_constraint,
Amar_constraint, Fay2007, Borowiec2006}. Possible constraints on
other types of Palatini-$f(R)$ model have also been considered
using big bang nucleosynthesis (BBN) and the requirements that the
success of the inflationary paradigm is not spoiled
\cite{Sotiriou_constraint}. Recently constraint from data on
matter power spectrum alone is given in \cite{Koivisto2006}. As
far as we know, there have been no attempts to confront Palatini
$f(R)$ gravity models with CMB data to date.

In this work, we will concentrate on the model of $f(R) = R +
\alpha (-R)^\beta$, where $\alpha$ is positive (so that it can
reproduce the recent cosmic acceleration). We refer to it as the
\emph{late} $f(R)$ cosmological model because its corrections to
GR dominate very lately, and we study \emph{both} its CMB
\emph{and} matter power spectra. For this, we need the
perturbation equations in the Palatini formalism, one set of which
has been worked out in \cite{K_perturb}. Here, however, we will
derive a set of covariant and gauge invariant perturbation
equations by the method of $3 + 1$ decomposition (see
Sec.~\ref{sect:Equations}) for our calculations. Also we shall try
to constrain the model parameters. Unlike previous works, we use
the full three-year WMAP data set \cite{WMAP3yr} instead of the
CMB shift parameter only. This $f(R)$ gravity model will be
constrained firstly by the WMAP CMB spectra data, and then jointly
by the CMB spectra, SNe measurements from Supernova Legacy Survey
(SNLS), plus the matter power spectrum data measured by the Sloan
Digital Sky Survey (SDSS) \cite{SDSS}.

The paper is organized as follows. In Sec.~\ref{sect:Equations},
we will first review briefly the theory of $ f(R)$ gravity in
the Palatini formalism, and we then present the perturbation
equations. Then in Sec.~\ref{sect:Constraints} the CMB and matter
power spectra for different choices of parameters are displayed and
discussed, and the constraints from various data sets will be
presented. Finally, we conclude in Sec.~\ref{sect:Conclusion}.
There is also an appendix where we show that our perturbation
equations are equivalent to those in the synchronous gauge under
specified conditions. Throughout this work we will assume a flat
universe filled with cold dark matter, photons, baryons, electrons
and 3 species of neutrinos (all massless); the unit $c = 1$ is
adopted. The metric convention used in this paper is $(+,-,-,-)$.

\section{Field Equations in Theories of Palatini - $f(R)$ Gravity}
\label{sect:Equations} In this section we shall first summarize
the properties of the general theory of $f(R)$ gravity in the
Palatini formalism, and we then derive its perturbation equations.
These equations could be found elsewhere \cite{Baojiu}, and here
we list them just for completeness.

\subsection{General theory of $f(R)$ gravity in the Palatini approach}

The starting point of our discussion on the Palatini-$f(R)$
gravity is the modified Einstein-Hilbert action, which is given as
\begin{eqnarray}
\label{eq:1} S &=& \int d^4
x\sqrt{-g}\left[\frac{1}{2\kappa}f(R)+\mathcal{L}_{m}\right],
\end{eqnarray}
where $\kappa = 8\pi G$ ($G$ is the Newton's constant) and
$R=g^{ab}R_{ab}(\bar{\Gamma})$ ($a, b = 0, 1, 2, 3$) with
$R_{ab}(\bar{\Gamma})$ being defined as
\begin{eqnarray}
\label{eq:2} R_{ab} &\equiv& \bar{\Gamma}^{c}_{ab, c} -
\bar{\Gamma}^{c}_{ac, b} +
\bar{\Gamma}^{c}_{cd}\bar{\Gamma}^{d}_{ab} -
\bar{\Gamma}^{c}_{ad}\bar{\Gamma}^{d}_{cb}\ .
\end{eqnarray}
Notice that the connection $\bar{\Gamma}$ here is not the
conventional Levi-Civita connection of the metric $g_{ab}$, which
we shall denote by $\Gamma$; rather it will be treated as an
independent field in the Palatini approach to the $f(R)$ theory of
gravity. Correspondingly, the tensor $R_{ab}$ and scalar $R$ are
also not the Ricci tensor and Ricci scalar calculated from
$g_{ab}$ as in GR, which instead are denoted by $\mathcal{R}_{ab}$
and $\mathcal{R}$ respectively in this work
($\mathcal{R}=g^{ab}\mathcal{R}_{ab}$). The matter Lagrangian
density $\mathcal{L}_{m}$, on the other hand, is assumed to be
independent of $\bar{\Gamma}$, which is the same as in GR.

The extremization of the action Eq.~(\ref{eq:1}) with respect to
the metric $g_{ab}$ then gives the modified Einstein equations
\begin{eqnarray}
\label{eq:3} FR_{ab} - \frac{1}{2}g_{ab}f(R) &=&
\kappa\mathcal{T}_{ab},
\end{eqnarray}
in which $F = F(R) \equiv \partial f(R)/\partial R$ and
$\mathcal{T}_{ab}$ is the energy-momentum tensor in the system
discussed. The trace of Eq.~(\ref{eq:3}) reads
\begin{eqnarray}
\label{eq:4} FR - 2f = \kappa\mathcal{T}
\end{eqnarray}
with $\mathcal{T}=\rho - 3p$ ($\rho$ is the energy density and $p$
the isotropic pressure) being the trace of the energy-momentum
tensor. This is the so-called structural equation
\cite{Allemandi2004a} which relates $R$ directly to the energy
components in the universe: given a specific form of $f(R)$ and
thus $F(R)$, $R$ can be obtained as a function of $\mathcal{T}$ by
numerically or analytically solving this equation.

The variation of Eq.~(\ref{eq:1}) with respect to the connection
field $\bar{\Gamma}$ leads to another equation
\begin{eqnarray}
\label{eq:5} \nabla_{a}[F(R)\sqrt{-g}g^{bc}] &=& 0,
\end{eqnarray}
which indicates that the connection $\bar{\Gamma}$ is compatible
with a metric $\gamma_{ab}$ that is conformal to $g_{ab}$:
\begin{eqnarray}
\label{eq:6} \gamma_{ab} &=& F(R)g_{ab}.
\end{eqnarray}
With the aid of Eq.~(\ref{eq:6}) we could now easily obtain the
relation between $R_{ab}$ and $\mathcal{R}_{ab}$ as
\begin{eqnarray}
\label{eq:7} R_{ab} &=& \mathcal{R}_{ab} +
\frac{3\mathcal{D}_{a}F\mathcal{D}_{b}F}{2F^{2}} -
\frac{\mathcal{D}_{a}\mathcal{D}_{b}F}{F} -
\frac{g_{ab}\mathcal{D}^{c}\mathcal{D}_{c}F}{2F}.
\end{eqnarray}
Note that in above we are using $\mathcal{D}$ and $\nabla$ to
denote the covariant derivative operators which are compatible
with $g_{ab}$ and $\gamma_{ab}$ respectively.

Since $\mathcal{L}_{m}$ depends only on $g_{ab}$ (and, of course,
some matter fields) and the energy-momentum  conservation law
holds with respect to it, we shall treat this metric as the
physical one. Consequently the difference between $f(R)$ gravity
and GR could be understood as a change in the manner in which the
spacetime curvature and thus the physical Ricci tensor
$\mathcal{R}_{ab}$ responds to the distribution of matter (through
the modified Einstein equations). In order to make this point
explicit, we can rewrite Eq.~(\ref{eq:3}) by the use of
Eq.~(\ref{eq:7}) as
\begin{eqnarray}
\label{eq:8}
\kappa\mathcal{T}_{ab} &=& F\mathcal{R}_{ab} -
\frac{1}{2}g_{ab}f\nonumber\\
&& + \frac{3}{2F}\mathcal{D}_{a}F\mathcal{D}_{b}F -
\mathcal{D}_{a}\mathcal{D}_{b}F -
\frac{1}{2}g_{ab}\mathcal{D}^{c}\mathcal{D}_{c}F,
\end{eqnarray}
in which $F(\mathcal{T}), f(\mathcal{T})$ are now simply functions
of $\mathcal{T}$.

\subsection{The Perturbation Equations}

The perturbation equations in general theories of $f(R)$ gravity
have been derived in \cite{K_perturb}. However, here we adopt a
different, covariant and gauge invariant derivation which utilizes
the method of $3+1$ decomposition \cite{Ellis1989, Challinor1999,
Lewis2000}.

The main idea of $3+1$ decomposition is to make space-time splits
of physical quantities with respect to the 4-velocity $u^{a}$ of
an observer. A projection tensor $h_{ab}$ is then defined as
$h_{ab}=g_{ab} - u_{a}u_{b}$ which could be used to obtain
covariant tensors orthogonal to $u$. For example, the covariant
spatial derivative $\hat{\mathcal{D}}$ of an arbitrary tensor
field $T^{b\cdot\cdot c}_{d\cdot\cdot e}$ (which, by definition,
is orthogonal to $u$) is given as
\begin{eqnarray}
\label{eq:9} \hat{\mathcal{D}}^{a}T^{b\cdot\cdot\cdot
c}_{d\cdot\cdot\cdot e} &\equiv&
h^{a}_{i}h^{b}_{j}\cdot\cdot\cdot\
h^{c}_{k}h^{r}_{d}\cdot\cdot\cdot\
h^{s}_{e}\mathcal{D}^{i}T^{j\cdot\cdot\cdot k}_{r\cdot\cdot\cdot
s}.
\end{eqnarray}
The energy-momentum tensor and covariant derivative of the
4-velocity $u$ could be decomposed respectively as
\begin{eqnarray}
\label{eq:10} \mathcal{T}_{ab} &=& \pi_{ab} + 2q_{(a}u_{b)} + \rho
u_{a}u_{b}
-ph_{ab},\\
\label{eq:11} \mathcal{D}_{a}u_{b} &=& \sigma_{ab} + \varpi_{ab} +
\frac{1}{3}\theta h_{ab} + u_{a}A_{b}.
\end{eqnarray}
In the above $\pi_{ab}$ is the projected symmetric trace free
(PSTF) anisotropic stress, $q$ the vector heat flux, $\sigma_{ab}$
the PSTF shear tensor, $\varpi_{ab} =
\hat{\mathcal{D}}_{[a}u_{b]}$, $\theta = \mathcal{D}^{a}u_{a} =
3\dot{a}/a$ ($a$ is the cosmic scale factor) the expansion scalar
and $A_{a} = \dot{u}_{a}$ the acceleration. The overdot expressed
as $\dot{\phi} =u^{a}\mathcal{D}_{a}\phi$ is the derivative with
respect to the proper time of the comoving observer moving at
velocity $u$, and the square brackets denote antisymmetrization
and parentheses symmetrization. The normalization is chosen to be
$u^{2}=1$ in consistence with our metric convention.

Decomposing the Riemann tensor and making use of the modified
Einstein equations with the general techniques used in GR, we
obtain, after linearization, five constraint equations
\begin{eqnarray}
\label{eq:12}
0 &=& \hat{\mathcal{D}}^{c}(\epsilon^{ab}_{\ \ cd}u^{d}\varpi_{ab});\\
\label{eq:13}
\frac{1}{F}\kappa q_{a} &=&
\frac{3\dot{F}\hat{\mathcal{D}}_{a}F}{2F^{2}} +
\frac{\theta\hat{\mathcal{D}}_{a}F}{3F} -
\frac{\hat{\mathcal{D}}_{a}\dot{F}}{F}\nonumber\\
&& -\frac{2}{3}\hat{\mathcal{D}}_{a}\theta +
\hat{\mathcal{D}}^{b}\sigma_{ab} +
\hat{\mathcal{D}}^{b}\varpi_{ab};\\
\label{eq:14}
\mathcal{B}_{ab} &=&
\left[\hat{\mathcal{D}}^{c}\sigma_{d(a}
+ \hat{\mathcal{D}}^{c}\varpi_{d(a}\right]\epsilon_{b)ec}^{\ \ \ \ d}u^{e};\\
\label{eq:15}
\hat{\mathcal{D}}^{b}\mathcal{E}_{ab} &=&
\frac{1}{2F}\kappa\left[\hat{\mathcal{D}}^{b}\pi_{ab} +
\left(\frac{2}{3}\theta + \frac{\dot{F}}{F}\right) q_{a} +
\frac{2}{3}\hat{\mathcal{D}}_{a}\rho\right]\nonumber\\
&& - \frac{1}{2F^{2}}\kappa(\rho + p)\hat{\mathcal{D}}_{a}F;\\
\label{eq:16}
\hat{\mathcal{D}}^{b}\mathcal{B}_{ab} &=&
\frac{1}{2F}\kappa\left[\hat{\mathcal{D}}_{c}q_{d} + (\rho +
p)\varpi_{cd}\right]\epsilon_{ab}^{\ \ cd}u^{b}.
\end{eqnarray}
Here $\epsilon_{abcd}$ is the covariant permutation tensor,
$\mathcal{E}_{ab}$ and $\mathcal{B}_{ab}$ are respectively the
electric and magnetic parts of the Weyl tensor
$\mathcal{W}_{abcd}$, given respectively by $\mathcal{E}_{ab} =
u^{c}u^{d}\mathcal{W}_{acbd}$ and $\mathcal{B}_{ab} = -
\frac{1}{2}u^{c}u^{d}\epsilon_{ac}^{\ \ ef}\mathcal{W}_{efbd}$.

We also obtain seven propagation equations:
\begin{eqnarray}
\label{eq:17}
\dot{\rho} + (\rho + p)\theta + \hat{\mathcal{D}}^{a}q_{a} &=& 0;\\
\label{eq:18}
\dot{q_{a}} + \frac{4}{3}\theta q_{a} + (\rho +
p)A_{a} - \hat{\mathcal{D}}_{a}p + \hat{\mathcal{D}}^{b}\pi_{ab} &=& 0;\\
\label{eq:19}
\dot{\theta} + \frac{1}{3}\left[\theta +
\frac{3\dot{F}}{2F}\right]\theta -
\hat{\mathcal{D}}^{a}A_{a}\nonumber\\
-\left[\frac{3\dot{F}^{2}}{2F^{2}} - \frac{3\ddot{F}}{2F} -
\frac{\kappa\rho}{F} - \frac{f}{2F} -
\frac{\hat{\mathcal{D}}^{2}F}{2F}\right] &=& 0;\\
\label{eq:20}
\dot{\sigma}_{ab}+ \frac{2}{3}\left[\theta +
\frac{3\dot{F}}{4F}\right]\sigma_{ab} - \hat{\mathcal{D}}_{\langle
a}A_{b\rangle}\nonumber\\
+ \mathcal{E}_{ab} + \frac{1}{2F}\kappa\pi_{ab} +
\frac{1}{2F}\hat{\mathcal{D}}_{\langle
a}\hat{\mathcal{D}}_{b\rangle}F &=& 0;\\
\label{eq:21}
\dot{\varpi} + \frac{2}{3}\theta\varpi -
\hat{\mathcal{D}}_{[a}A_{b]} &=& 0;\\
\label{eq:22}
\frac{1}{2F}\kappa\left[\dot{\pi}_{ab} +
\left(\frac{1}{3}\theta -
\frac{3\dot{F}}{2F}\right)\pi_{ab}\right]\nonumber\\ -
\frac{1}{2F}\kappa\left[(\rho + p)\sigma_{ab} \ +
\hat{\mathcal{D}}_{\langle a}q_{b\rangle}\right]\nonumber\\
-\left[\dot{\mathcal{E}}_{ab} + \left(\theta +
\frac{\dot{F}}{2F}\right)\mathcal{E}_{ab} -
\hat{\mathcal{D}}^{c}\mathcal{B}_{d(a}\epsilon_{b)ec}^{\ \ \ \ d}u^{e}\right] &=& 0;\\
\label{eq:23}
\dot{\mathcal{B}}_{ab} + \left(\theta +
\frac{\dot{F}}{2F}\right)\mathcal{B}_{ab} +
\hat{\mathcal{D}}^{c}\mathcal{E}_{d(a}
\epsilon_{b)ec}^{\ \ \ \ d}u^{e} \nonumber\\
+ \frac{1}{2F}\kappa\hat{\mathcal{D}}^{c}
\mathcal{\pi}_{d(a}\epsilon_{b)ec}^{\ \ \ \ d}u^{e} &=& 0,
\end{eqnarray}
The angle brackets mean taking the trace free part of a quantity.

Besides the above equations, it would also be useful to express
the projected Ricci scalar $\hat{\mathcal{R}}$ in the
hypersurfaces orthogonal to $u^{a}$ (the projected Riemann tensor,
$\hat{\mathcal{R}}_{abcd}$, is defined by $[\hat{\mathcal{D}}_{a},
\hat{\mathcal{D}}_{b}]v^{c} = \hat{\mathcal{R}}_{abd}^{\ \ \ \
c}v^{d}$, similar to the definition of the full covariant Riemann
tensor $\mathcal{R}_{abcd}$ but with a conventional opposite sign,
and the calculations for the projected Ricci tensor
$\hat{\mathcal{R}}_{ab}$ and projected Ricci scalar
$\hat{\mathcal{R}}$ just follow the same way as in GR) as
\begin{eqnarray}
\label{eq:24} \hat{\mathcal{R}} &\doteq& \frac{\kappa(\rho +
3p)-f}{F} - \frac{2}{3}\left[\theta +
\frac{3\dot{F}}{2F}\right]^{2} -
\frac{2\hat{\mathcal{D}}^{2}F}{F}.
\end{eqnarray}
The spatial derivative of this projected Ricci scalar, $\eta_{a}
\equiv \frac{1}{2}a\hat{\mathcal{D}}_{a}\hat{\mathcal{R}}$, is
then obtained after a lengthy calculation as
\begin{eqnarray}
\label{eq:25}
\eta_{a} &=&
\frac{a}{2F}\kappa(\hat{\mathcal{D}}_{a}\rho +
3\hat{\mathcal{D}}_{a}p) - \frac{a}{F}\left[\frac{3}{2F}\dot{F} +
\theta\right]\hat{\mathcal{D}}_{a}\dot{F}\nonumber\\
&& - \frac{a}{2F}\hat{\mathcal{D}}_{a}f -
\frac{a}{F}\hat{\mathcal{D}}_{a}(\hat{\mathcal{D}}^{2}F) -
\frac{2a}{3}\left[\frac{3}{2F}\dot{F} +
\theta\right]\hat{\mathcal{D}_{a}}\theta\nonumber\\
&& + \frac{a}{3F}\left[\frac{3}{2F}\dot{F} +
\theta\right]\left[\frac{3}{2F}\dot{F} -
\theta\right]\hat{\mathcal{D}}_{a}F,
\end{eqnarray}
and its time evolution is governed by the propagation equation
\begin{eqnarray}
\label{eq:26}
\dot{\eta}_{a} + \frac{2\theta}{3}\eta_{a} &=&
\frac{a}{2F}\left[\frac{3}{F}\dot{F} -
\frac{2}{3}\theta\right]\hat{\mathcal{D}}_{a}\hat{\mathcal{D}}^{2}F
-
\frac{a}{F}\kappa\hat{\mathcal{D}}_{a}\hat{\mathcal{D}}^{c}q_{c}\nonumber\\
&& -
\frac{a}{F}\hat{\mathcal{D}}_{a}(\hat{\mathcal{D}}^{2}F)^{\cdot} -
\left[\frac{\dot{F}}{F} +
\frac{2\theta}{3}\right]a\hat{\mathcal{D}}_{a}\hat{\mathcal{D}}^{c}A_{c}.\
\ \ \ \
\end{eqnarray}

As we are considering a spatially flat universe, its spatial
curvature will vanish for large scales, meaning that
$\hat{\mathcal{R}}=0$. Thus from Eq.~(\ref{eq:24}) we have
\begin{eqnarray}
\label{eq:27} \left[\frac{1}{3}\theta + \frac{\dot{F}}{2F}\right]
^{2} &=& \frac{1}{6F}\left[\kappa(\rho+3p)-f\right].
\end{eqnarray}
This is just the modified (first) Friedmann equation in the $f(R)$
version of gravitational theory, and the other modified background
equations (the second Friedmann equation and the
energy-conservation equation) could be obtained by taking the
zeroth-order parts of Eqs.~(\ref{eq:17}) and (\ref{eq:19}). It is
easy to check that when $f(R) = R$, we have $F = 1$, and these
equations just reduce to those in GR -- in this case GR and the
Palatini-$f(R)$ theory lead to the same results. In the appendix
we also show that this set of perturbation equations is equivalent
to that derived in the synchronous gauge, which serves as
a  check for this work.

Remember that we have had $f, F$ and $R$ as functions of
$\mathcal{T}$ at hand, it is then straightforward to calculate
$\dot{F}, \ddot{F}, \hat{\mathcal{D}}_{a}F,
\hat{\mathcal{D}}_{a}\dot{F}$ \emph{etc.} as functions of
$\dot{\mathcal{T}} \doteq - (\rho_{b} + \rho_{c})\theta$ and
$\hat{\mathcal{D}}_{a}\mathcal{T} =
(1-3c_{s}^{2})\hat{\mathcal{D}}_{a}\rho_{b} +
\hat{\mathcal{D}}_{a}\rho_{c}$, in which $\rho_{b(c)}$ is the
energy density of baryons (cold dark matter) and $c_{s}$ is the
baryon sound speed. Note that in this work we choose to neglect
the small baryon pressure except in the terms where its spatial
derivative is involved, in which case they might be significant at
small scales. The above equations could then be numerically
propagated given the initial conditions, to obtain  the evolutions
of small density perturbations and the CMB and matter power
spectra in theories of $f(R)$ gravity. Finally the three-year WMAP
data on CMB spectra, SNLS SN data and SDSS data on matter power
spectrum could be used to constrain parameters in the $f(R)$
models. These results will be given in the following section.

\section{NUMERICAL RESULTS AND COSMOLOGICAL CONSTRAINTS ON THE MODEL}
\label{sect:Constraints}

This section is devoted to numerical results and constraints of
the present model. To this effect we will first very briefly
summarize and explain the effects of the $f(R)$ modifications to
GR on the linear spectra; for more details see \cite{Baojiu}.
After that we shall employ the public Markov Chain Monte Carlo
(MCMC) engine \cite{Lewis2002} to search the parameter space with
the theoretical CMB and matter power spectra calculated by the
modified CAMB code; the constraints are then summarized and
discussed.

\begin{figure}
  \centering
  \includegraphics[scale=0.85] {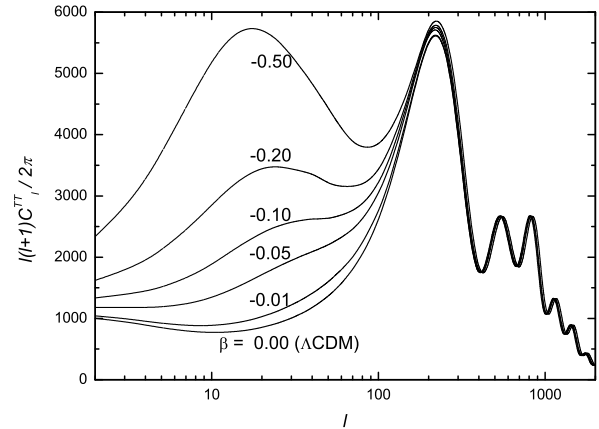}
  \caption{The TT CMB spectrum for the $f(R) = R + \alpha(-R)^\beta$ model,
  with $\Omega_m$ (current fractional energy density of nonrelativistic matter)
  and $H_{0}$ (current Hubble constant) fixed to be 0.3 and
  72 km/s/Mpc respectively. Choices of $\beta$ are indicated besides the
  curves. The case $\beta = 0$ corresponds to a $\Lambda\mathrm{CDM}$ Universe.}
  \label{fig:Figure1}
\end{figure}

In Fig.~\ref{fig:Figure1} we have displayed the TT CMB spectra for
the model with different choices of $\beta$. It is obvious from
this figure that, when $\beta < 0$, the spectrum gets a boost in
the scales $l \leq 100$, which could be significant if $|\beta|$
is large enough. This effect is due to a strong late-time
integrated Sachs-Wolfe (ISW) effect \cite{Baojiu, Koivisto2005},
which in turn originates from the unusually rapid late-time decay
of the gravitatoinal potential $\phi$ of the present $f(R)$ model
compared with $\Lambda\mathrm{CDM}$, as shown in the lower panel
of Fig.~3 in \cite{Baojiu} (see this reference for more details).
We have not given the curves for $\beta > 0$ because in that case
the spectrum generally blows up except for very small $|\beta|$s
(see below).

Another interesting feature in Fig.~\ref{fig:Figure1} is that for
negative $\beta$ the spectrum shifts towards the right-hand-side
(larger $l$'s), likely due to the unusual
angular-distance-redshift relation \cite{Riazuelo2002, Uzan2003}.
Since the standard $\Lambda\mathrm{CDM}$ cosmology is expected to
be valid in the early times when the correction to GR is
negligible, the sound horizon  and the thickness of the last
scattering surface are the same as in $\Lambda\mathrm{CDM}$. But
at late times the Friedmann equation is modified
(Eq.~(\ref{eq:27})), and so is the relation between redshift and
conformal distance. This would cause the CMB spectrum to shift
sideways. This shifting effect, however, is negligible for the
constrained ranges of $\beta$ obtained below.

The CMB EE polarization and cross correlation spectra show no
additional interesting features and cannot be used to put strong
constraint on the model parameters, and so we will not present and
discuss them here.

\begin{figure}
  \centering
  \includegraphics[scale=0.85] {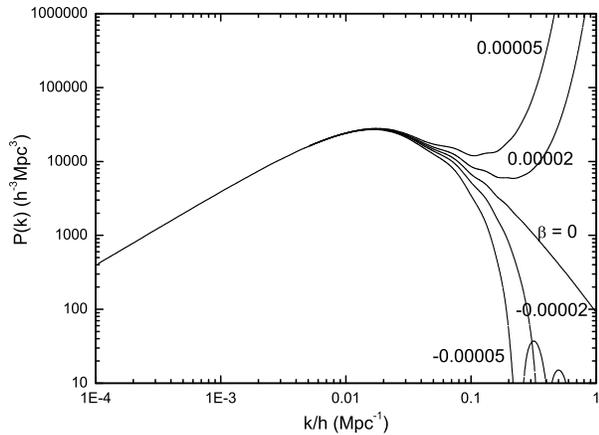}
  \caption{The matter power spectra of the $f(R) = R + \alpha(-R)^\beta$ model
  for different choices of $\beta$ (both negative and positive) as indicated beside
  the curves. The case $\beta = 0$ corresponds to a $\Lambda\mathrm{CDM}$ Universe.}
  \label{fig:Figure2}
\end{figure}

We have also given in Fig.~\ref{fig:Figure2} the matter power
spectra. As indicated in this figure, the matter power spectrum
depends sensitively on the value of $\beta$ and could differ from
$\Lambda\mathrm{CDM}$ significantly even if $|\beta|$ only
deviates from 0 by a tiny amount \emph{e.g.}, of order
$\mathcal{O}(10^{-5})$. This feature has been pointed out and
discussed extensively in \cite{K_perturb, Koivisto2006, Baojiu}.
Basically, this is because of the sensitive response of the
modified gravity to the spatial variations of matter distribution,
which, at small enough scales would significantly affect the
growth of density perturbations. To be explicit, for large enough
$k$'s, the growth equation for the comoving energy density
fluctuations $\delta_m$ could be written as \cite{Koivisto2006}
\begin{eqnarray}
\label{eq:28} \frac{d^{2}\delta_m}{dN^{2}} &\doteq& -
\frac{k^{2}}{a^{2}H^{2}}\frac{\dot{F}}{3F(2FH+\dot{F})}\delta_m,
\end{eqnarray}
in which $N \equiv \log(a)$ and
$\dot{F}/3F(2FH+\dot{F})=c_{s,eff}^{2}$ acts as an effective sound
speed squared that vanishes in the $\Lambda\mathrm{CDM}$ model.
For the present $f(R)$ model, we have $F = 1 - \alpha\beta(-
R)^{\beta - 1}$, in which $\alpha$ and $-R$ are positive and $- R$
decreases with time. So if $\beta < 0$, then  $\dot{F} > 0$ and
thus $c_{s,eff}^{2} > 0$; this effective pressure term will
restrict the growths of small-scale density perturbations and
leads to oscillations (as shown in Fig.~\ref{fig:Figure2} for the
case of $\beta = -0.00005$) of the spectrum in these scales. On
the other hand, if $\beta > 0$ (as we hope to recover standard
$\Lambda\mathrm{CDM}$ cosmology in earlier times, we shall also
restrict ourselves to $\beta < 1$), then $\dot{F}$ and
$c_{s,eff}^{2}$ will be negative; this will make the density
fluctuations unstable and blow up, the same reason why the CMB
spectrum depends so sensitively on positive $\beta$s.

Since from Figs.~\ref{fig:Figure1} and \ref{fig:Figure2} we have
seen that the linear spectra of our $f(R)$ model depend very
sensitively on the model parameter $\beta$, it can be expected
that the data on CMB and matter power spectra could place
stringent constraints on $\beta$, as we will show now. As
mentioned in Sec.~\ref{sect:Intro}, we shall firstly use the full
three year WMAP data and then perform a joint constraint
simultaneously using WMAP, SNLS and SDSS data to constrain the
parameters.

\begin{figure}
  \centering
  \includegraphics[scale=0.48] {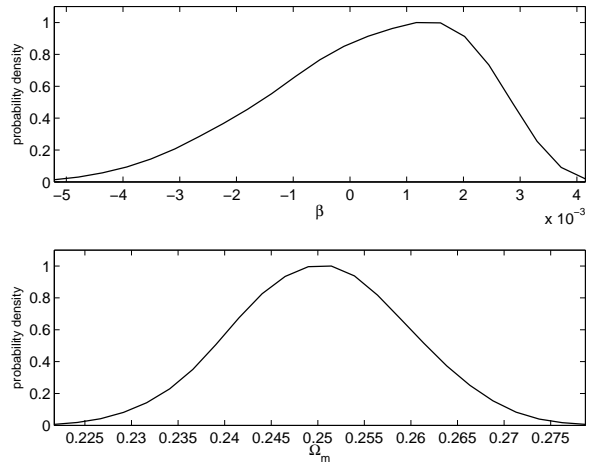}
  \caption{The marginal distributions of $\Omega_m$ and $\beta$,
  obtained using the three year WMAP data alone. Here the distributions
  are normalized such that the maximum probability density is 1.}
  \label{fig:Figure3}
\end{figure}

\begin{figure}
  \centering
  \includegraphics[scale=0.48] {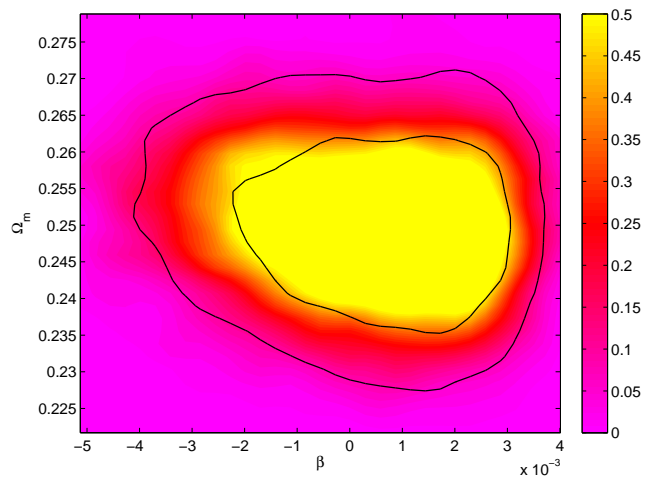}
  \caption{(Color Online) The contour plot of the joint distribution of $\Omega_m$
  and $\beta$, constrained by WMAP data alone. The inner and outer
  loops are the 68\% and 95\% confidence contours respectively.}
  \label{fig:Figure4}
\end{figure}

Because the Hubble parameter $H_0$ is already measured to rather
good precision by the HST Key Project, we shall use $H_0=72\
\mathrm{km/s/Mpc}$ \cite{Freedman2001} in our calculations.
Therefore we vary the following parameters: baryon density
$\omega_b = \Omega_b h^2$, cold dark matter $\omega_c = \Omega_c
h^2 $, reionization redshift $z_{re}$, spectral index $n_s$,
normalization amplitude $A_s$ and the model parameter $\beta$. In
Fig.~\ref{fig:Figure3} the marginal distributions of $\Omega_m$
and $\beta$ are shown. The 95\% confidence interval for $\Omega_m$
and $\beta$ are $[0.233, 0.268]$ and $[-3.45\times 10^{-3}, 3.07
\times 10^{-3}]$ respectively. We also present the contour plot of
the joint distribution of $\Omega_m$ and $\beta$ in
Fig.~\ref{fig:Figure4}. From these figures we can see that the CMB
spectra could constrain $|\beta|$ to $\mathcal{O}(10^{-3})$, $\sim
100$ times more stringent than the constraint from the CMB shift
parameter \cite{Amar_constraint}, which is of order 0.1. That the
CMB spectra is much more powerful in constraining the parameters
than the CMB shift parameter is expected because the former bears
a lot more information than the latter. It looks from these
figures that a slightly positive $\beta$ is preferred by the CMB
data.

\begin{figure*}
  \centering
  \includegraphics[scale=0.85] {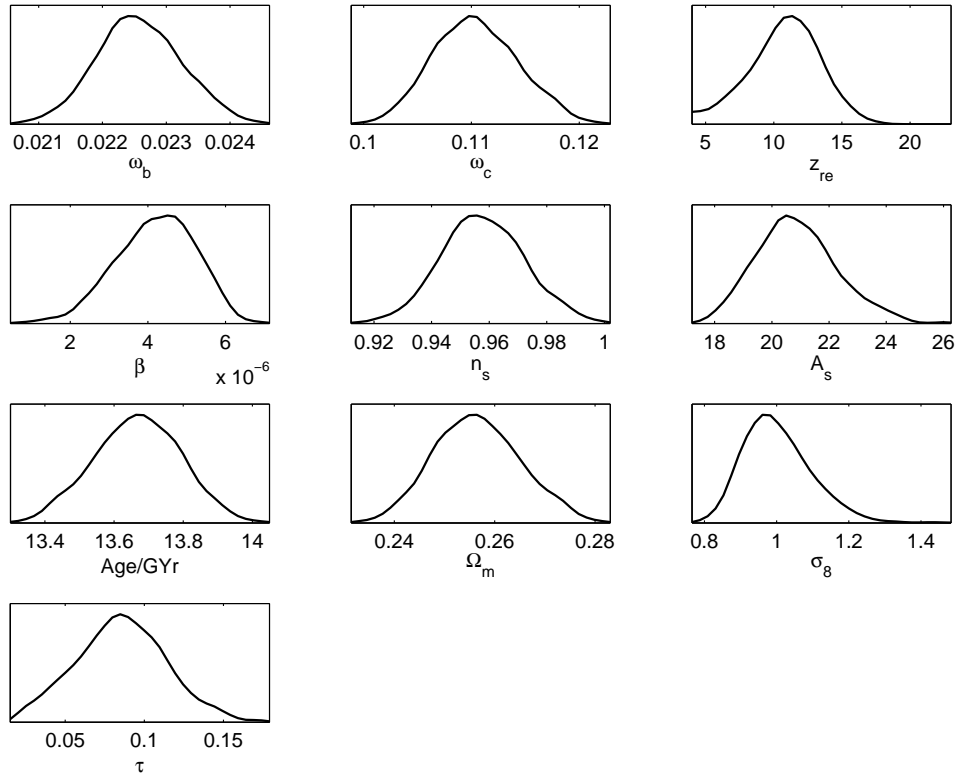}
  \caption{The marginal distributions of the various model parameters,
  constrained simutaneously by the WMAP, SNLS and SDSS data sets. The distributions
  are normalized such that the maximum probability density is 1.}
  \label{fig:Figure5}
\end{figure*}

\begin{figure}
  \centering
  \includegraphics[scale=0.48] {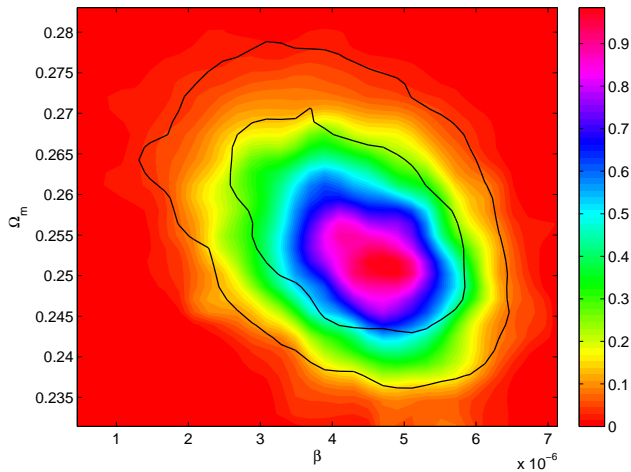}
  \caption{(Color Online) The contour plot of joint distribution of $\Omega_m$
  and $\beta$ under the constraints of the WMAP, SNLS and SDSS data sets.
  The inner and outer loops are the 68\% and 95\% confidence
  contours respectively.}
  \label{fig:Figure6}
\end{figure}

\begin{figure}
  \centering
  \includegraphics[scale=1.2] {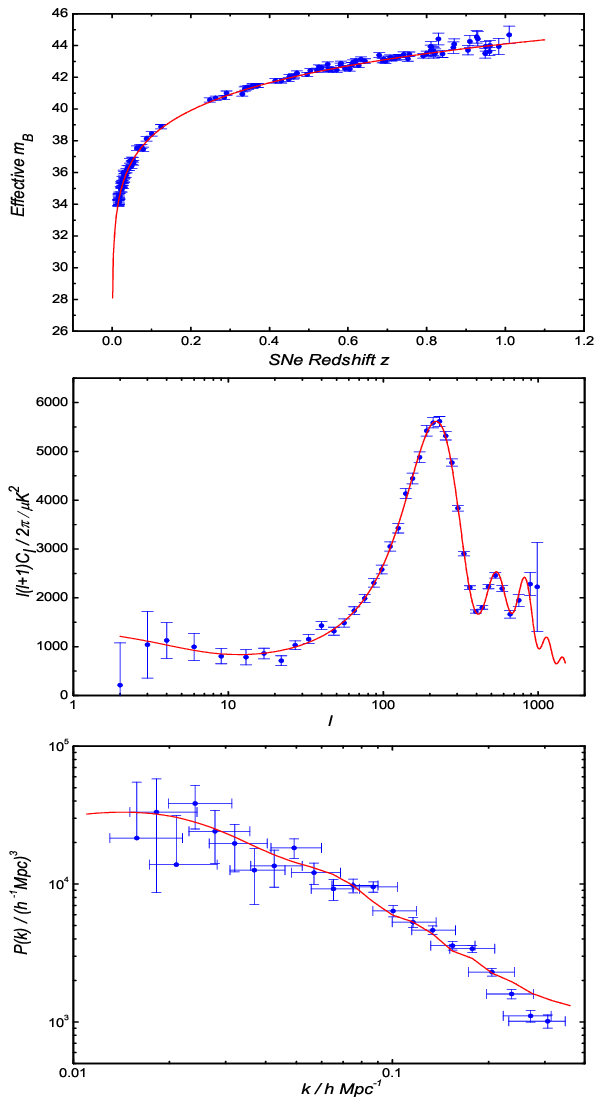}
  \caption{(Color Online) The data points from SNLS (upper panel), WMAP
  (middle panel) and SDSS (lower panel) data sets against the theoretical
  curves of our best-fitted model. For SDSS data we have plotted both
  vertical and horizonal error bars. Note that the last three data points
  from SDSS (for which $k > 0.2h\ \mathrm{Mpc}^{-1}$) are not included in
  our numerical constraints, and we have used a bias of 1.1 to relate the best 
  fit theoretical power spectrum to the SDSS data.}
  \label{fig:Figure7}
\end{figure}

To tighten the bounds on the parameters, we perform a joint
constraint making use of both the three-year WMAP and the SDSS
data sets. In addition to the matter power spectrum, the SNe data
from SNLS \cite{Astier2006} is also used in the joint constraint,
though their effects are found to be negligible. For the SDSS
data, we conservatively adopt the measurements for scales larger
than $k = 0.2 h\ \mathrm{Mpc}^{-1}$ (where $h \equiv H_{0}/(100\
\mathrm{km\ s^{-1} Mpc^{-1}})$) to avoid encountering the
nonlinear effects in the measured matter power spectrum. The bias
between galaxy power spectrum and matter power spectrum is assumed
to be a scale independent constant; CosmoMC \cite{Lewis2002}
assumes a flat prior on it and marginalizes analytically.

The calculation indicates that indeed the allowed range is shrunk,
as indicated in Figs.~\ref{fig:Figure5} and \ref{fig:Figure6} (for
completeness in Fig.~\ref{fig:Figure7} we have also plotted the
best-fitted curves with the observational data points from SNLS,
WMAP and SDSS we use in the constraints). The 95\% confidence
intervals for $\Omega_m $ and $\beta$ now become [0.241, 0.274]
and $[2.12 \times 10^{-6}, 5.98 \times 10^{-6}]$ respectively, and
the 95\% confidence interval for $\sigma_{8}$ is $[0.85, 1.21]$.
The distribution of $\Omega_m$ does not change much since it is
already well constrained by the WMAP, but the bound on $\beta$ is
tightened to the order of $10^{-6}$ (and obviously future refined
data could still further this constraint). What is more, these
joint constraints also prefer a positive $\beta$ and actually have
excluded the case of $\beta = 0$, \emph{i.e.}, the
$\Lambda\mathrm{CDM}$ paradigm, at the 95\% confidence level. In
Fig.~\ref{fig:Figure7} we could see that all the data sets are
fitted very well. Furthermore, more stringent constraints on
$\beta$ can be obtained if data points in the nonlinear regime are
also used since the matter power spectrum will blow up in the
small scales for positive $\beta$ (see Fig.~\ref{fig:Figure2}).
Nonetheless, our stringent constraint on $\beta$ already makes the
model nearly indistinguishable from $\Lambda\mathrm{CDM}$ for
practical purposes, and without a natural motivation for such tiny
values of $\beta$ this model should be more reasonably disfavored.

\

\section{Discussion and Conclusion}
\label{sect:Conclusion}

In conclusion, we have in this work derived the perturbation
equations for general theories of Palatini-$f(R)$ gravity and
applied them to a typical class of model $f(R) = R + \alpha
(-R)^\beta$ which is proposed as an alternative to the
cosmological constant to account for the late-time accelerating
cosmic expansion and has been extensively studied. We then
calculate the CMB and matter power spectra for this model using a
modified CAMB code. It is shown that for negative $\beta$s the
potential $\phi$ will see an unusually rapid decay at late times,
leading to an enhancement of the ISW effect and thus a boost of
the TT CMB spectrum at small $l$s. There also appears a positive
effective pressure term in the equation governing the growth of
density perturbations, which could be significant for small scales
(large $k$s) and restricts the perturbation growths in these
scales. For positive $\beta$s, however, the small-scale density
fluctuations will become unstable and grow exponentially,
resulting in blowing-ups of the matter power spectrum.

We have constrained the model parameters using the
WMAP, SNLS and SDSS data. Because the CMB and matter power spectra
are rather sensitive to the exact values of the parameter $\beta$,
we are able to give much more stringent constraints on $\beta$
($\mathcal{O}(10^{-3})$ and $\mathcal{O}(10^{-6})$ respectively)
than those ($\mathcal{O}(10^{-1})$) coming from the CMB shift
parameter fitting or measurements on SNIa \cite{Amar_constraint}.
Compared with the bound ($\mathcal{O}(10^{-5})$) from SDSS data
alone \cite{Koivisto2006}, our WMAP + SNLS + SDSS constraint is
tighter because the allowed range of $\Omega_m$ is largely reduced
here.

These constraints seem to make the present model (in its allowed
parameter space) indistinguishable from the $\Lambda\mathrm{CDM}$
paradigm and raise a fine-tuning problem to the late $f(R)$
gravity theory. However, there still remains the interesting
possibility that $f(R)$ modification of gravity enters at earlier
times (or higher densities): can it survive the tests from WMAP
and SDSS data? This topic is beyond the scope of this article and
has been investigated in another work \cite{Baojiu}; in that case
the parameter space for $f(R)$ gravity is also highly limited.

\section*{Appendix}

This appendix is devoted to showing the equivalence between the
covariant perturbation equations derived here and the perturbation
equations in the synchronous gauge \cite{Ma1995}. In the latter
case, the line element is expressed as (indices $i, j, k$ run over
1, 2, 3 hereafter)
\begin{eqnarray}
\label{eq:29}
ds^{2} &=& a(\tau)^{2}\left[d\tau^{2} - (\delta_{ij}
+ h^{S}_{ij}) dx^{i}dx^{j}\right],
\end{eqnarray}
in which $\tau$ is the conformal time given by $dt = a(\tau)d\tau$
and we have defined the scalar modes in Fourier space as
\begin{widetext}
\begin{eqnarray}
\label{eq:30}
h^{S}_{ij}(\mathbf{x}, \tau) &=& \int d^{3}k\
\exp(i\mathbf{k}\cdot\mathbf{x})\left[\hat{k}_{i}\hat{k}_{j}h^{S}(\mathbf{k},
\tau) + 6\left(\hat{k}_{i}\hat{k}_{j} -
\frac{1}{3}\delta_{ij}\right)\eta^{S}(\mathbf{k}, \tau)\right],
\end{eqnarray}
\end{widetext}
where a superscript $'S'$ denotes quantities in the synchronous
gauge and $\hat{k} = \mathbf{k}/k$ is the unit vector in the
$\mathbf{k}$-direction.

In order to relate the synchronous gauge variables $h^{S}$ and
$\eta^{S}$ with those in the covariant perturbation method, let us
do the harmonic expansion for the first-order quantities $h_{a}
\equiv \hat{\mathcal{D}}_{a}a$ and $\eta_{a}$ as $h_{a} =
\sum_{k}khQ_{a}^{k}$ and $\eta_{a} = \sum_{k}
\frac{k^{3}}{a^{2}}\eta Q_{a}^{k}$ (Here $Q_{a}^{k} =
\frac{a}{k}\hat{\mathcal{D}}_{a}Q^{k}$ and $Q^{k}$ is the
eigenfunction of the generalized Helmholtz equation
$a^{2}\hat{\mathcal{D}}^{2}Q^{k} = k^{2}Q^{k}$. For more details
see \emph{e.g.,} \cite{Challinor1999, Lewis2000}), to obtain the
variables $h$ and $\eta$ relevant to specified $k$-modes. We shall
choose the reference velocity $u$ to be the 4-velocity of cold
dark matter, which means that the cold dark matter heat flux $q_c
= \rho_c v_c$ and the acceleration $A$ both vanish
\cite{Challinor1999, Lewis2000}. Then using the relations
$\eta^{S} = - \eta/2$ and $h'^{S} = 6h'$ \cite{Lewis2005} (a prime
here means taking derivative with respect to the conformal time
$\tau$) it is easy to show the equivalence between these two sets
of perturbation equations. More explicitly: Eq.~(\ref{eq:26}) is
equal to the following first-order equation
\begin{widetext}
\begin{eqnarray}
\label{eq:31}
\bar{F}k^{2}\eta'^{S} &\doteq&
\frac{\kappa}{2}a^{2}(\bar{\rho}^{S} + \bar{p}^{S})\theta^{S} -
\frac{1}{2}\left[\frac{3\bar{F}'}{2\bar{F}} +
3\mathcal{H}\right]k^{2}\delta F + \frac{1}{2}k^{2}\delta F',
\end{eqnarray}
\end{widetext}
where $\mathcal{H} \equiv a'/a$. Taking the spatial derivative of
Eq.~(\ref{eq:19}) leads to
\begin{widetext}
\begin{eqnarray}
\label{eq:32}
-\frac{\bar{F}}{2}\left[h''^{S} + \mathcal{H}h'^{S}
+ \frac{\bar{F}'}{2\bar{F}}h'^{S}\right] &\doteq& \kappa
a^{2}\delta\rho + \frac{a^{2}}{2}\frac{\partial
f(\bar{T})}{\partial \bar{T}}\delta T -
\frac{3\bar{F}'}{\bar{F}}\delta F' + \frac{3\delta F''}{2} +
\left[\frac{3a''}{a} - 3\mathcal{H}^{2} +
\frac{3}{2}\left(\frac{\bar{F}'}{\bar{F}}\right)^{2} +
\frac{k^2}{2}\right]\delta F.
\end{eqnarray}
\end{widetext}
Then from Eqs.~(\ref{eq:25}) and (\ref{eq:32}) we can obtain
\begin{widetext}
\begin{eqnarray}
\label{eq:33}
\bar{F}\left[4k^{2}\eta^{S} -
\frac{5\mathcal{H}h'^{S}}{2} - \frac{h''^{S}}{2} -
\frac{5\bar{F}'}{4\bar{F}}h'^{S}\right] &\doteq&
\frac{3a^{2}}{2}\frac{\partial f(\bar{T})}{\partial \bar{T}}\delta
T - \kappa a^{2}\delta T_{i}^{i} + 6\mathcal{H}\delta F' +
\frac{3\delta F''}{2} + \left[\frac{3a''}{a} + 3\mathcal{H}^2 +
\frac{5k^2}{2}\right]\delta F,
\end{eqnarray}
\end{widetext}
and finally, taking time derivative of Eq.~(\ref{eq:26}) and
making use of Eqs.~(\ref{eq:19}), (\ref{eq:25}), (\ref{eq:27}) and
(\ref{eq:32}), we get
\begin{widetext}
\begin{eqnarray}
\label{eq:34} \bar{F}\left[\frac{2}{3}k^{2}\eta -
\frac{2}{3}\mathcal{H}\left(h'^{S} + 6\eta'^{S}\right) -
\frac{1}{3}\left(h''^{S} + 6\eta''^{S}\right)\right] &\doteq&
\kappa a^{2}(\bar{\rho} + \bar{p})\sigma^{S} +
\frac{2}{3}k^{2}\delta F + \frac{1}{2}\bar{F}'\left(h'^{S} +
6\eta'^{S}\right).
\end{eqnarray}
\end{widetext}
In the above $\delta$ means the spatial variation of
a quantity for the discussed length scale (or $k$) and a bar its
background value (up to first order all the barred quantities in
these equations could be equally replaced by their unbarred
correspondences). The variables $\theta^{S}, \sigma^{S}$ are
defined respectively as \cite{Ma1995} $(\bar{\rho} +
\bar{p})\theta^{S} \equiv ik^{j}\delta T^{0}_{j}$ and $(\bar{\rho}
+ \bar{p})\sigma^{S} \equiv -
(\hat{k}^{j}\hat{k}_{i}-\frac{1}{3}\delta^{j}_{i})(T^{i}_{j} -
\frac{1}{3}\delta^{i}_{j}T^{k}_{k})$. $T_{ab}$ is the
energy-momentum tensor in the synchronous gauge (to be
distinguished from the $\mathcal{T}_{ab}$ used above) and $T$ its
trace.

We have checked that Eqs.~(\ref{eq:31}) - (\ref{eq:34}) are just
the perturbation equations one has in the synchronous gauge for
theories of Palatini-$f(R)$ gravity, as expected
\cite{Challinor1999, Lewis2000}.

\begin{acknowledgments}
We would like to thank the ITSC of the Chinese University of Hong
Kong for using their clusters for numerical calculations. The work
is partially supported by a grant from the Research Grants Council
of the Hong Kong Special Administrative Region, China (Project
No.~400707). B.~L.~acknowledges supports from the Overseas
Research Studentship, Cambridge Overseas Trust and the Department
of Applied Mathematics and Theoretical Physics at the University
of Cambridge. We are grateful to Dr. Tomi Koivisto for his helpful
discussions.
\end{acknowledgments}

\appendix

\newcommand{\noopsort}[1]{} \newcommand{\printfirst}[2]{#1}
  \newcommand{\singleletter}[1]{#1} \newcommand{\switchargs}[2]{#2#1}

\end{document}